\documentclass[11pt]{article}
\usepackage{appb,epsfig}

\def\Journal#1#2#3#4{{#1} {\bf #2}, #3 (#4)}

\def\AP{\em Ann. Phys.}

\def\NPA{{\em Nucl. Phys.} A}
\def\NPB{{\em Nucl. Phys.} B}
\def\PLB{{\em Phys. Lett.}  B}

\def\PRL{\em Phys. Rev. Lett.}
\def\PRC{{\em Phys. Rev.} C}
\def\PRD{{\em Phys. Rev.} D}
\def\ZP{\em Z. Phys.}

\def\be{\begin{equation}}
\def\ee{\end{equation}}
\def\bea{\begin{eqnarray}}
\def\eea{\end{eqnarray}}

\begin{document}
\title{Meson-nucleon scattering and vector mesons\\ in nuclear
matter\thanks{Invited talk presented at the Meson98 workshop,
Cracow, 30.5.-2.6.98} }
\author{Bengt Friman
\address{Gesellschaft f\"ur Schwerionenforschung (GSI)\\
D-64220 Darmstadt, Germany\\
\&\\Institut f\"ur Kernphysik, TU Darmstadt\\
D-64289 Darmstadt, Germany }}

\maketitle

\begin{abstract}
The properties of vector mesons in nuclear matter are discussed. I
examine the constraints imposed by elementary processes on the
widths of $\rho$ and $\omega$ mesons in nuclear matter.
Furthermore, results for the $\rho$- and $\omega$-nucleon
scattering amplitudes obtained by fitting meson-nucleon scattering
data in a coupled-channel approach are presented.
\end{abstract}

\section{Introduction}
The electromagnetic decay of the vector mesons into $e^+e^-$ and
$\mu^+\mu^-$ pairs makes them particularly well suited for
exploring the conditions in dense and hot matter in nuclear
collisions. The lepton pairs provide virtually undistorted
information on the mass distribution of the vector mesons in the
medium.

The lepton-pair spectrum in nucleus-nucleus collisions at SPS
energies exhibits a low-mass enhancement compared to proton-proton
and proton-nucleus collisions \cite{Helios}. A quantitative
interpretation of the lepton-pair data can be obtained within a
scenario, where the effective vector-meson masses are reduced in a
hadronic environment \cite{likobr,cassing,BR,Hatsuda-Lee}. On the
other hand, attempts to interpret the low-mass enhancement of
lepton pairs in terms of many-body effects also yield good
agreement with the data \cite{CRW,CBRW}. In these calculations the
broadening of the $\rho$ meson in nuclear matter due to the
interactions of its pion cloud with the medium \cite{HFN,CS,AK,KKW}
and the momentum dependence of the $\rho$-meson self energy due to
the coupling with baryon-resonance--nucleon-hole states
\cite{FP,PLPM} are taken into account.

Through the low-density theorem hadron-nucleon scattering data can
be used to determine the self-energies of hadrons in nuclear matter
at low densities. In this talk I discuss the the constraints on the
imaginary part of the vector-meson self energy in nuclear matter
that can be derived from elementary reactions, and present results
of a coupled channel calculation of meson-nucleon scattering. The
latter provides a model for the vector-meson--nucleon scattering
amplitude.

\section{Constraints from elementary processes}

The low-density theorem states that the self energy of e.g. a
vector meson $V$ in nuclear matter is given by~\cite{LDT}
\begin{equation}
\Sigma_V(\rho_N) = -4\pi(1+\frac{m_V}{m_N})\langle{f}_{V N}\rangle\,\, \rho_N
+ \dots,
\label{LDT}
\end{equation}
where $m_V$ is the mass of the vector meson, $m_N$ that of the
nucleon, $\rho_N$ the nucleon density and $\langle{f}_{V N}\rangle$
denotes the $V N$ forward scattering amplitude $f_{V N}$,
appropriately averaged over the nucleon Fermi sea. For the vector
mesons $\rho, \omega$ and $\phi$ the elastic scattering amplitudes
have to be extracted indirectly, e.g. in a coupled channel
approach, which I will discuss in section 3.

In order to avoid an extrapolation over a wide range in mass, which
would introduce a strong model dependence~\cite{seoul}, I will use
only data in the relevant kinematic range to constrain the $V N$
scattering amplitudes. As an example, I shall first discuss the
implications of the data on pion-induced vector-meson production
for the in-medium width of $\omega$ mesons.

Using detailed balance and unitarity one can relate the cross
section for the reaction $\pi^- p \rightarrow \omega n$ to the
imaginary part of the $\omega$-nucleon scattering amplitude due to
the $\pi^- p$ channel~\cite{seoul}
\begin{figure}[t]
\center{\epsfig{file=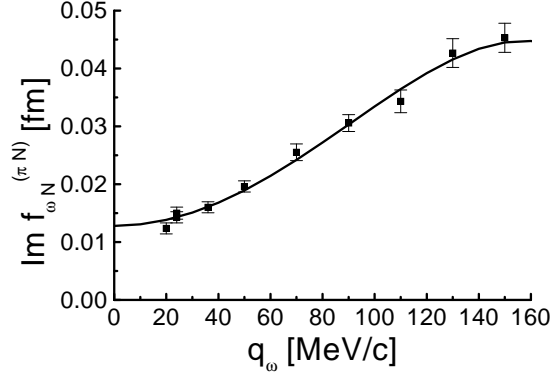,height=55mm}}
\caption{\label{omefit} The $\pi^- p$ contribution to the
imaginary part of the $\omega -n$ forward-scattering amplitude
obtained by a fit to the $\pi^- p \rightarrow \omega n$ data
(ref.~\cite{keyne-karami}) near threshold.}
\end{figure}
\begin{equation}
\sigma_{\pi^- p \rightarrow \omega n} =
12 \pi \frac{k_\omega}{k_\pi^2}\, \mbox{Im}\,\bar{f}_{\omega
n\rightarrow\omega n}^{\,(\pi^- p)}(\theta=0),
\label{pi-ome}
\end{equation}
where $\bar{f}$ denotes the spin-averaged scattering amplitude.
Close to the $\omega n$ threshold, the scattering amplitude can be
expanded in powers of the relative momentum in the $\omega n$
channel $q_\omega$. An excellent fit to the data from threshold up
to $q=120$ MeV/c$^2$ is obtained with $\mbox{Im}\,\bar{f}_{\omega
n\rightarrow\omega n}^{\,(\pi^- p)}=a + b q_\omega^2 + c
q_\omega^4$, where $a = 0.013$ fm, $b = 0.10$ fm$^3$ and $c = -
0.08$ fm$^5$ (see Fig.~\ref{omefit}). The coefficient $a$ is the
imaginary part of the scattering length.

The corresponding contribution of the $\pi$-nucleon channel to the
width of the $\omega$ meson at rest in nuclear matter can now be
obtained by using the low-density theorem~(\ref{LDT})
\begin{equation}
\Delta\Gamma_\omega = 4\pi(1+\frac{m_\omega}{m_N})\frac{3}{2}
\frac{\langle \mbox{Im}\,
f_{\omega n\rightarrow\omega n}^{\,(\pi^- p)}\rangle
\,\rho_N}{m_\omega}.
\label{gam-ome}
\end{equation}
This implies that at nuclear-matter density the width of the
$\omega$ meson in nuclear matter is increased by 9 MeV due to the
$\pi$-nucleon channel. Other channels, like the $\pi\pi N$ channel
leads to a further enhancement of the $\omega$ width in matter.

For the $\rho$ meson the situation is more complicated. First of
all the experimentally accessible $\pi N$ channel is subdominant.
Second, both isospin 1/2 and 3/2 are allowed. Thus, three
independent reactions are needed to pin down the amplitudes of the
two isospin channels and their relative phase. The
data~\cite{rhoprod,lanbor} on the reactions $\pi^-
p\rightarrow\rho^0 n$, $\pi^+ p\rightarrow \rho^+ p$ and $\pi^-
p\rightarrow \rho^- p$ would, if measured down to threshold, be
sufficient to determine the amplitudes. Unfortunately the large
width of the $\rho$ meson makes its identification close to
threshold very difficult. Thus, the data even in the one channel,
which is measured close to threshold $\pi^- p\rightarrow \rho^0 n$,
is afflicted with a large uncertainty~\cite{manley}. Clearly new
data on $\rho$ production close to threshold would be very useful.

\section{Meson-nucleon scattering}

In this section I describe a coupled channel approach to
meson-nucleon scattering~\cite{FLW}. The following channels are
included: $\pi N$, $\rho N$, $\omega N$, $\pi \Delta$ and $\eta N$.
Our goal is to determine the vector-meson--nucleon scattering
amplitude close to threshold, which in turn determines the self
energy of a vector-meson at rest in nuclear matter to leading order
in density.
\begin{figure}[t]
\center{\epsfig{file=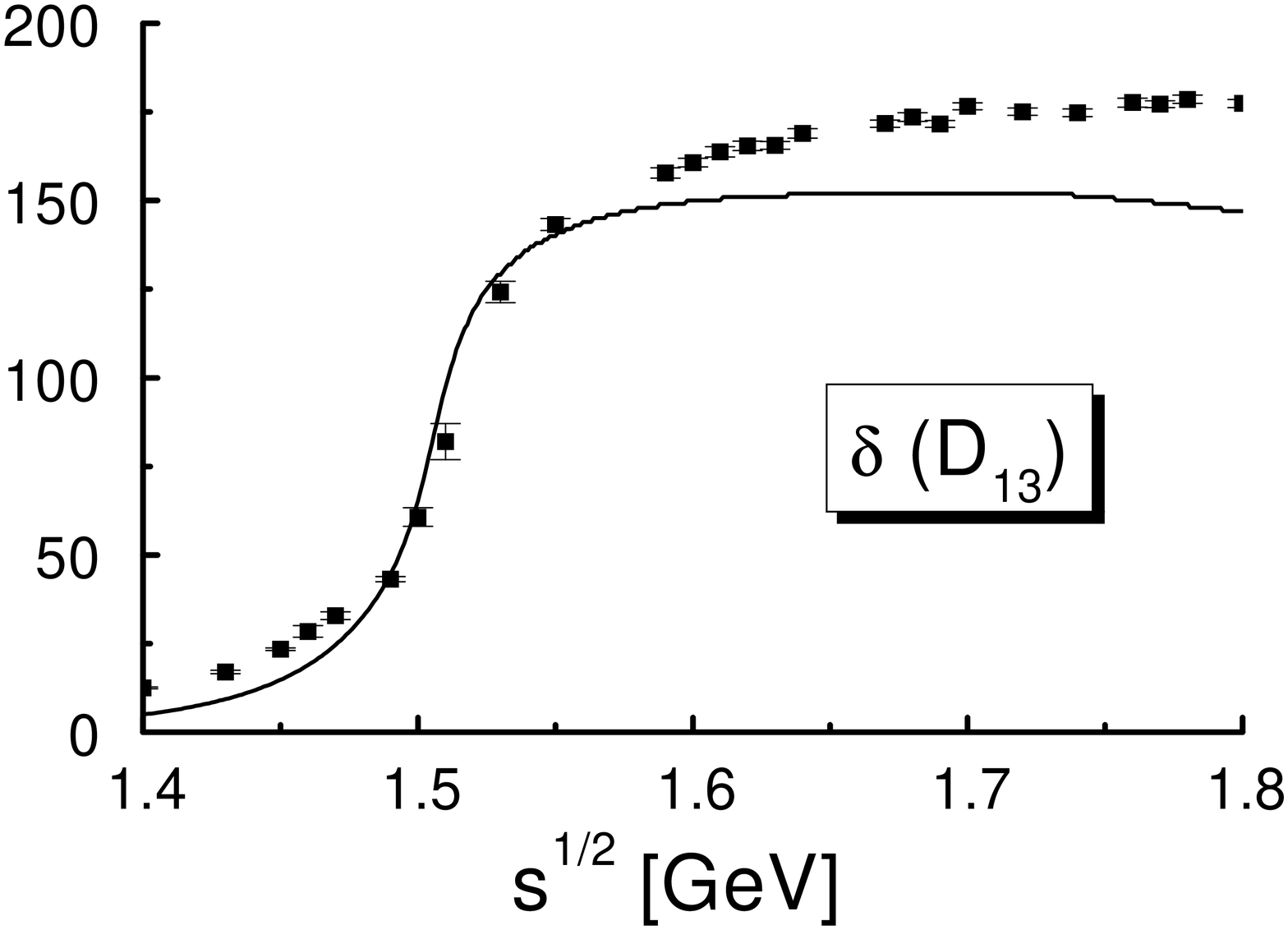,height=55mm}}
\center{\epsfig{file=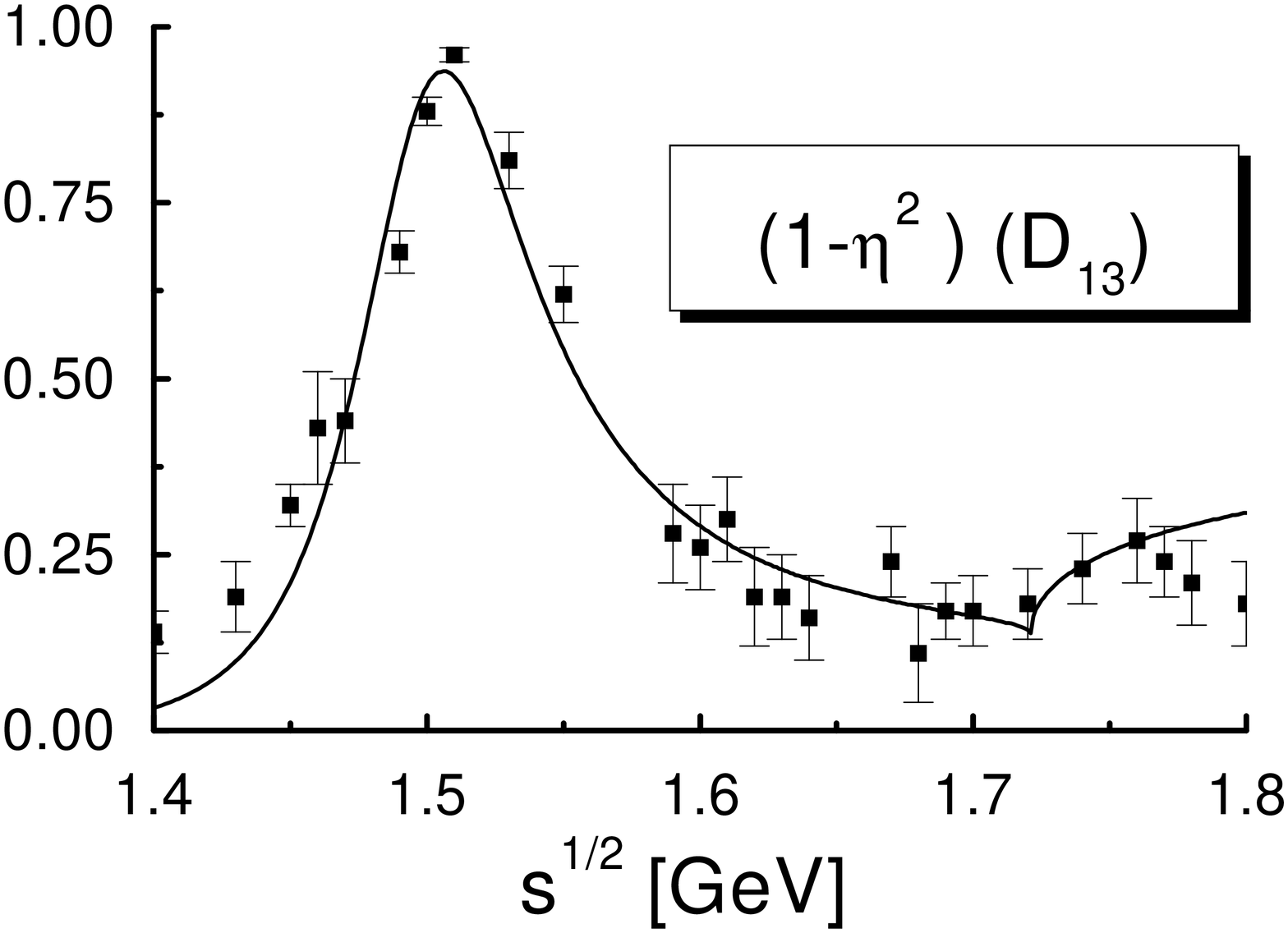,height=55mm}}
\caption{\label{D13-scatt} The $\pi N$ scattering phase shifts
and inelasticity. The line shows the best fit, while the data
points are those of the analysis of Arndt {\em et
al.}~\cite{arndt}.}
\end{figure}

Since we are interested in the vector-meson scattering amplitude
close to threshold, it is sufficient to consider only s-wave
scattering in the $\rho N$ and $\omega N$ channels. This implies
that in the $\pi N$ and $\pi\Delta$ channels we need only s- and
d-waves. In particular, we consider the $S_{11}, S_{31}, D_{13}$
and $D_{33}$ partial waves of $\pi N$ scattering. Furthermore, we
consider the pion-induced production of $\eta$, $\omega$ and $\rho$
mesons off nucleons. In order to learn something about the momentum
dependence of the vector-meson self energy, vector-meson--nucleon
scattering also in higher partial waves would have to be
considered.

In accordance with the approach outlined above only data in the
relevant kinematical range will be used in the analysis. The
threshold for vector-meson production off a nucleon is at $\sqrt{s}
\simeq 1.7$ GeV. We fit the data in the energy range
$1.45$ GeV $\leq \sqrt{s} \leq 1.8$ GeV, with an effective
Lagrangian with 4-point meson-meson--baryon-baryon interactions.
For details the reader is referred to ref.~\cite{FLW}. In
Fig.~\ref{D13-scatt} our fit to the $\pi N$ scattering data is
illustrated by the $D_{13}$ channel. In the remaining channels the
quality of the fit is in general better.
\begin{figure}[t]
\center{\epsfig{file=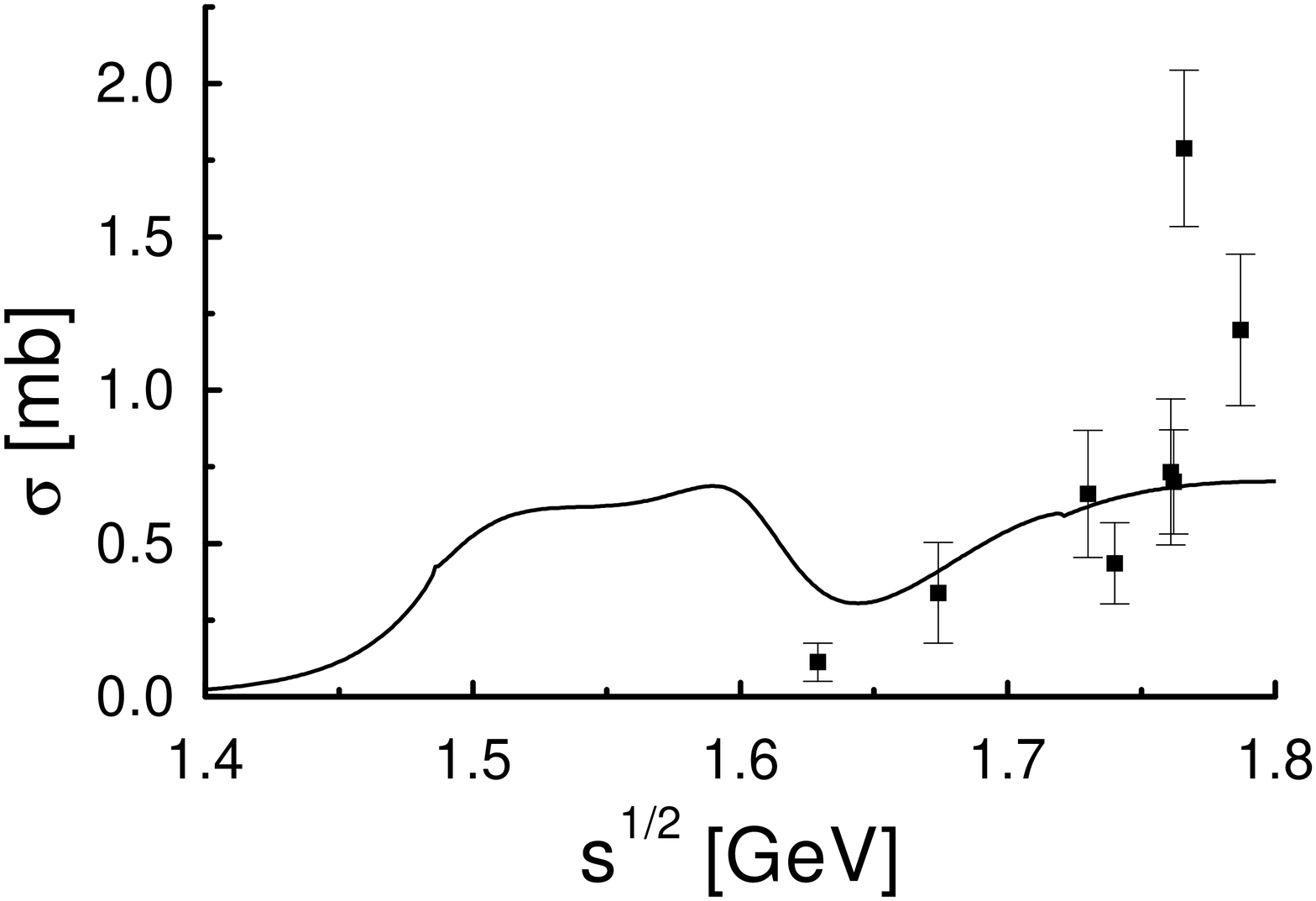,height=55mm}}
\caption{\label{rho-prod} The cross section for the reaction
$\pi^- p \rightarrow \rho^0 n$. The data points are from
ref.~\cite{rhoprod}, as given in ref.~\cite{lanbor}}
\end{figure}

Furthermore, in Fig.~\ref{rho-prod} the cross section for the
reaction $\pi^- p\rightarrow \rho^0 n$ is shown. The bumps at
$s^{1/2}$ below 1.6 GeV are due to the coupling to resonances below
the threshold, like the $N^\star(1520)$. This indicates that these
resonances may play an important role in the $\rho$-nucleon
dynamics, in agreement with the results of Manley and
Saleski~\cite{manley}. However, the strength of the coupling is
uncetrain, due to the ambiguity in the $\rho$-production cross
section close to threshold mentioned above. We find that also the
$\omega$ meson couples strongly to these resonances.

The pion-induced $\eta$-production cross section is well described
up to $s^{1/2} \simeq 1.65$ GeV. At higher energies presumably
higher partial waves, not included in our model, become important.
Similarly, the pion-induced production of $\omega$ mesons is
reasonably well represented close to threshold, although the strong
energy dependence of the amplitude shown in Fig.~\ref{omefit} is
not reproduced by the model. This may be due to the coupling to
channels not included at present, like the $K-\Sigma$ channel.
\begin{figure}[t]
\center{\epsfig{file=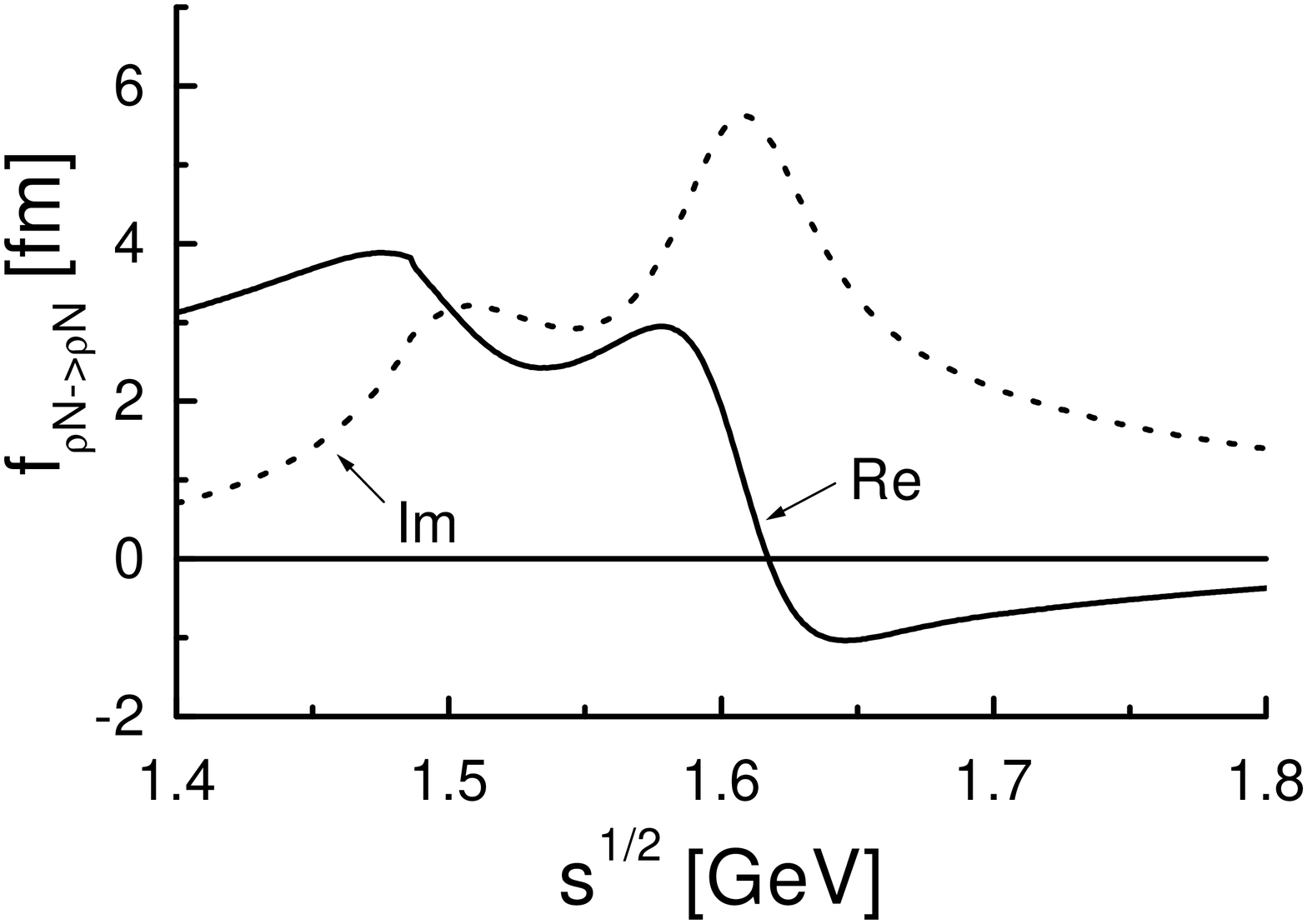,height=55mm}}
\center{\epsfig{file=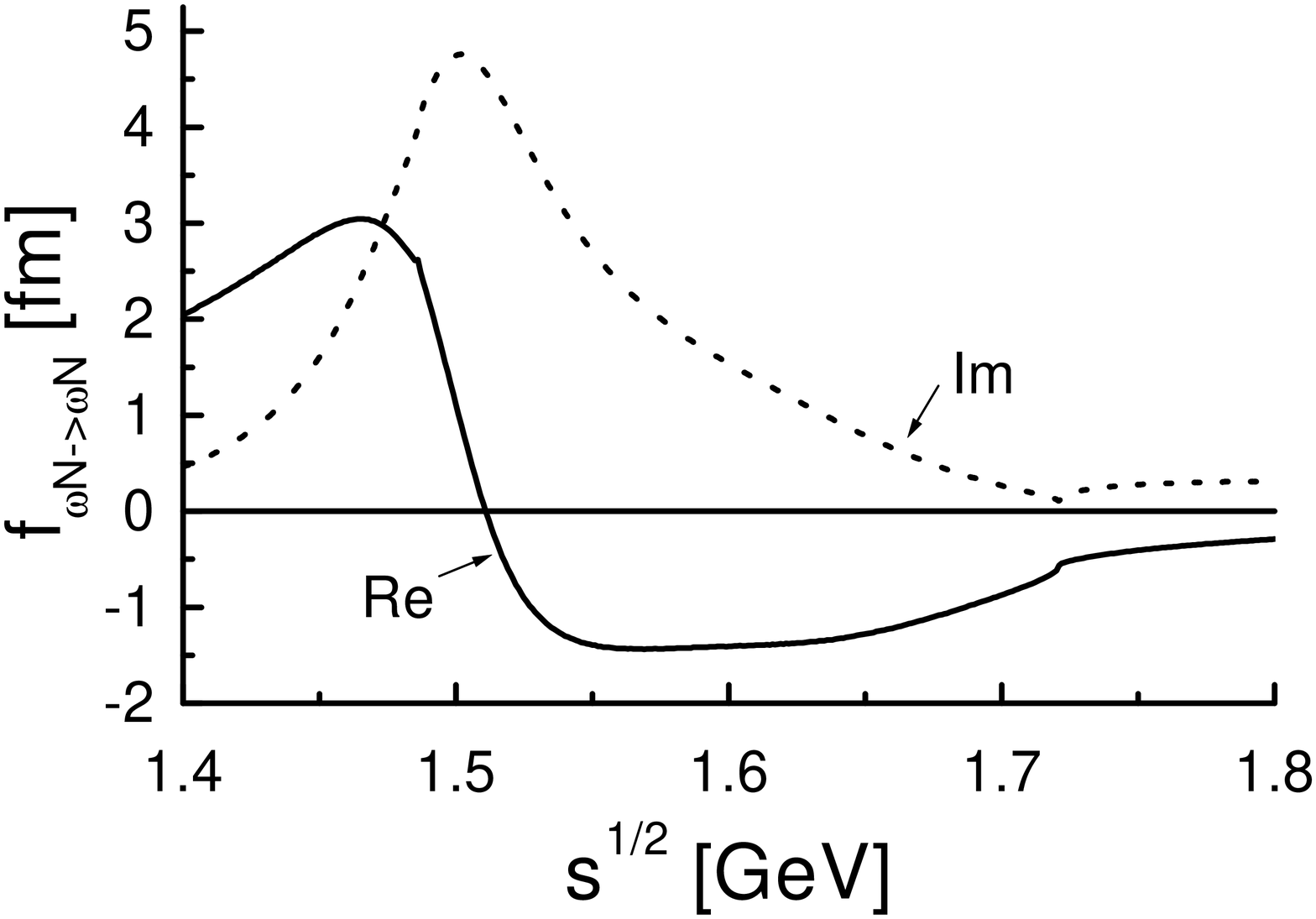,height=55mm}}
\caption{\label{amplitudes} The $\rho N$ and $\omega N$ scattering
scattering amplitudes, averaged over spin and isospin.}
\end{figure}

The resulting $\rho$- and $\omega$-nucleon scattering amplitudes
are shown in Fig.~\ref{amplitudes}. The $\rho-N$ and $\omega-N$
scattering lengths are (-0.2+0.7i) fm and (-0.5+0.1i) fm
respectively. To lowest order in the density, this corresponds to
the following in-medium modifications of masses and widths at
nuclear matter density: $\Delta m_\rho \simeq 20$ MeV, $\Delta
m_\omega \simeq 50$ MeV, $\Delta\Gamma_\rho \simeq 140$ MeV and
$\Delta \Gamma_\omega \simeq 20$ MeV. However, the coupling of the
vector mesons to baryon resonances below threshold, which is
reflected in the strong energy dependence of the amplitudes,
implies that in medium the vector-meson strength will be split into
a meson like mode, which is pushed up in energy, and a
resonance-hole like mode, which is pushed down in energy. The
downward shift of vector-meson strength would contribute to the
low-mass enhancement in the lepton-pair spectra. Thus, our results
support the dynamical scenario discussed in ref.~\cite{brown}.

\section*{Acknowledgments}

I thank M.~Lutz and G.~Wolf for valuable discussions. I am grateful
to W.~Florkowski and M.~Nowak for their warm hospitality during my
stay in Cracow and I acknowledge the partial support by the Polish
Government Project (KBN) grant 2P03B00814 as well as by the
Institute of Nuclear Physics in Cracow.

\end{document}